%% file: main.tex
\pdfoutput=1
\documentclass[12pt,a4paper]{article}

\usepackage{ifthen} 
\usepackage{subcaption}
\usepackage{comment}
\newboolean{pdflatex}
\setboolean{pdflatex}{true} 

\newboolean{articletitles}
\setboolean{articletitles}{true} 

\newboolean{uprightparticles}
\setboolean{uprightparticles}{false} 

\def\paperauthors{LHCb collaboration} 
\def\paperasciititle{Search for the lepton-flavour
violating decays B+ -> pi+ mu e} 
\def\papertitle{Search for the lepton-flavour
violating decays \Btopimue} 
\def\paperkeywords{{High Energy Physics}, {LHCb}} 
\def\papercopyright{\the\year\ CERN for the benefit of the LHCb collaboration} 
\def\paperlicence{CC BY 4.0 licence}
\def\paperlicenceurl{https://creativecommons.org/licenses/by/4.0/}

\newif\ifEnableSectionTOCLinks
\EnableSectionTOCLinksfalse 

\input{preamble}
\usepackage{longtable} 

\begin{document}

\renewcommand{\thefootnote}{\fnsymbol{footnote}}
\setcounter{footnote}{1}

\input{title-LHCb-PAPER}


\renewcommand{\thefootnote}{\arabic{footnote}}
\setcounter{footnote}{0}


\cleardoublepage


\pagestyle{plain} 
\setcounter{page}{1}
\pagenumbering{arabic}


\input{body}


\input{acknowledgements}

\input{End_Matter}




\clearpage

\addcontentsline{toc}{section}{References}
\bibliographystyle{LHCb}
\bibliography{main,standard,LHCb-PAPER,LHCb-CONF,LHCb-DP,LHCb-TDR}

\newpage
\input{Authorship_LHCb-PAPER-2026-013}

\end{document}

%% file: preamble.tex
\usepackage[top=1in, bottom=1.25in, left=1in, right=1in]{geometry}

\usepackage{microtype}
\usepackage{lineno}  
\usepackage{xspace} 
\usepackage{caption} 

\usepackage{graphicx}  
\usepackage{color}
\usepackage{colortbl}
\graphicspath{{./figs/}} 

\usepackage{amsmath} 
\usepackage{amssymb}
\usepackage{amsfonts}
\usepackage{upgreek} 

\newcommand*\patchAmsMathEnvironmentForLineno[1]{%
\expandafter\let\csname old#1\expandafter\endcsname\csname #1\endcsname
\expandafter\let\csname oldend#1\expandafter\endcsname\csname
end#1\endcsname
 \renewenvironment{#1}%
   {\linenomath\csname old#1\endcsname}%
   {\csname oldend#1\endcsname\endlinenomath}%
}
\newcommand*\patchBothAmsMathEnvironmentsForLineno[1]{%
  \patchAmsMathEnvironmentForLineno{#1}%
  \patchAmsMathEnvironmentForLineno{#1*}%
}
\AtBeginDocument{%
\patchBothAmsMathEnvironmentsForLineno{equation}%
\patchBothAmsMathEnvironmentsForLineno{align}%
\patchBothAmsMathEnvironmentsForLineno{flalign}%
\patchBothAmsMathEnvironmentsForLineno{alignat}%
\patchBothAmsMathEnvironmentsForLineno{gather}%
\patchBothAmsMathEnvironmentsForLineno{multline}%
\patchBothAmsMathEnvironmentsForLineno{eqnarray}%
}

\usepackage[pdftex,
            pdfauthor={\paperauthors},
            pdftitle={\paperasciititle},
            pdfkeywords={\paperkeywords}]{hyperref}
\usepackage{hyperxmp}
\hypersetup{
    pdfcopyright={Copyright (C) \papercopyright},
    pdflicenseurl={\paperlicenceurl}
}

\usepackage[colorinlistoftodos,textsize=scriptsize]{todonotes}

\usepackage[bottom,flushmargin,hang,multiple]{footmisc}

\usepackage[all]{hypcap} 

\input{lhcb-symbols-def} 

\input{local-symbols-def} 

\hypersetup{
  colorlinks   = true, 
  urlcolor     = blue, 
  linkcolor    = blue, 
  citecolor    = red   
}

\ifEnableSectionTOCLinks
    \usepackage[explicit]{titlesec} 
    
    \let\oldcontentsline\contentsline
    \renewcommand

    \titleformat{\section}{\normalfont\Large\bf}{\hyperlink{tocsection.\thesection}{{\thesection} \parbox[t]{\dimexpr\textwidth-1pc}{#1}}}{1pc}{}

    \titleformat{\subsection}{\normalfont\bf}{\hyperlink{tocsubsection.\thesubsection}{{\thesubsection} \parbox[t]{\dimexpr\textwidth-1pc}{#1}}}{1pc}{}

    \titleformat{name=\section,numberless}[display]{}{}{0pt}{\normalfont\Huge\bfseries #1}
\fi

\usepackage{cite} 
\usepackage{mciteplus}

%% file: lhcb-symbols-def.tex
\usepackage{xspace}
\usepackage{upgreek}

\def\lhcb   {\mbox{LHCb}\xspace}

\def\babar  {\mbox{BaBar}\xspace}

\def\cleo   {\mbox{CLEO}\xspace}

\def\MagUp {\mbox{\em Mag\kern -0.05em Up}\xspace}

\ifthenelse{\boolean{uprightparticles}}%
{

 \def\Pmu         {\ensuremath{\upmu}\xspace}
 \def\Pnu         {\ensuremath{\upnu}\xspace}
 
 \def\Ppi         {\ensuremath{\uppi}\xspace}

 \def\Ppsi        {\ensuremath{\uppsi}\xspace}

 \def\PDelta      {\ensuremath{\Delta}\xspace}
 \def\PXi         {\ensuremath{\Xi}\xspace}
 \def\PLambda     {\ensuremath{\Lambda}\xspace}
 \def\PSigma      {\ensuremath{\Sigma}\xspace}
 \def\POmega      {\ensuremath{\Omega}\xspace}
 \def\PUpsilon    {\ensuremath{\Upsilon}\xspace}
 \let\oldPi\Pi
 \def\PPi         {\ensuremath{\oldPi}\xspace}

 \def\PB      {\ensuremath{\mathrm{B}}\xspace}
 \def\PD      {\ensuremath{\mathrm{D}}\xspace}
 \def\PJ      {\ensuremath{\mathrm{J}}\xspace}
 \def\PK      {\ensuremath{\mathrm{K}}\xspace}
 \def\Pb      {\ensuremath{\mathrm{b}}\xspace}
 \def\Pc      {\ensuremath{\mathrm{c}}\xspace}
 \def\Pd      {\ensuremath{\mathrm{d}}\xspace}
 \def\Pe      {\ensuremath{\mathrm{e}}\xspace}
 \def\Pp      {\ensuremath{\mathrm{p}}\xspace}

 \def\Ps      {\ensuremath{\mathrm{s}}\xspace}

 \def\thebaroffset{0.0em}
}
{

 \def\Pmu         {\ensuremath{\mu}\xspace}
 \def\Pnu         {\ensuremath{\nu}\xspace}
 
 \def\Ppi         {\ensuremath{\pi}\xspace}

 \def\Ppsi        {\ensuremath{\psi}\xspace}
 
 \mathchardef\PDelta="7101
 \mathchardef\PXi="7104
 \mathchardef\PLambda="7103
 \mathchardef\PSigma="7106
 \mathchardef\POmega="710A
 \mathchardef\PUpsilon="7107
 \mathchardef\PPi="7105
 \def\PB      {\ensuremath{B}\xspace}
 \def\PD      {\ensuremath{D}\xspace}
 \def\PJ      {\ensuremath{J}\xspace}
 \def\PK      {\ensuremath{K}\xspace}
 \def\Pb      {\ensuremath{b}\xspace}
 \def\Pc      {\ensuremath{c}\xspace}
 \def\Pd      {\ensuremath{d}\xspace}
 \def\Pe      {\ensuremath{e}\xspace}
 \def\Pp      {\ensuremath{p}\xspace}

 \def\Ps      {\ensuremath{s}\xspace}

 \def\thebaroffset{0.18em}
}
\newcommand{\offsetoverline}[2][\thebaroffset]{\kern #1\overline{\kern -#1 #2}}%

\makeatletter
\ifcase \@ptsize \relax
  \newcommand{\miniscule}{\@setfontsize\miniscule{4}{5}}
\or
  \newcommand{\miniscule}{\@setfontsize\miniscule{5}{6}}
\or
  \newcommand{\miniscule}{\@setfontsize\miniscule{5}{6}}
\fi
\makeatother

\DeclareRobustCommand{\optbar}[1]{\shortstack{{\miniscule (\rule[.5ex]{1.25em}{.18mm})}
  \\ [-.7ex] $#1$}}

\def\en         {{\ensuremath{\Pe^-}}\xspace}   
\def\ep         {{\ensuremath{\Pe^+}}\xspace}

\def\emp        {{\ensuremath{\Pe^\mp}}\xspace}

\def\mup        {{\ensuremath{\Pmu^+}}\xspace}
\def\mun        {{\ensuremath{\Pmu^-}}\xspace} 
\def\mupm       {{\ensuremath{\Pmu^\pm}}\xspace}

\def\mumu       {{\ensuremath{\Pmu^+\Pmu^-}}\xspace}

\def\ellm       {{\ensuremath{\ell^-}}\xspace}
\def\ellp       {{\ensuremath{\ell^+}}\xspace}
\def\ellpm      {{\ensuremath{\ell^\pm}}\xspace}

\def\neu        {{\ensuremath{\Pnu}}\xspace}
\def\neub       {{\ensuremath{\overline{\Pnu}}}\xspace}
\def\neue       {{\ensuremath{\neu_e}}\xspace}

\def\neumb      {{\ensuremath{\neub_\mu}}\xspace}
\def\neul       {{\ensuremath{\neu_\ell}}\xspace}
\def\neulb      {{\ensuremath{\neub_\ell}}\xspace}

\def\dquark    {{\ensuremath{\Pd}}\xspace}

\def\squark    {{\ensuremath{\Ps}}\xspace}

\def\cquark    {{\ensuremath{\Pc}}\xspace}

\def\bquark    {{\ensuremath{\Pb}}\xspace}

\def\pion   {{\ensuremath{\Ppi}}\xspace}
\def\piz    {{\ensuremath{\pion^0}}\xspace}
\def\pip    {{\ensuremath{\pion^+}}\xspace}
\def\pim    {{\ensuremath{\pion^-}}\xspace}

\def\kaon    {{\ensuremath{\PK}}\xspace}

\def\KorKbar {\kern \thebaroffset\optbar{\kern -\thebaroffset \PK}{}\xspace}

\def\Kp      {{\ensuremath{\kaon^+}}\xspace}
\def\Km      {{\ensuremath{\kaon^-}}\xspace}

\def\Kstarp  {{\ensuremath{\kaon^{*+}}}\xspace}

\def\Dbar    {{\ensuremath{\offsetoverline{\PD}}}\xspace}
\def\D       {{\ensuremath{\PD}}\xspace}

\def\DorDbar {\kern \thebaroffset\optbar{\kern -\thebaroffset \PD}\xspace}
\def\Dz      {{\ensuremath{\D^0}}\xspace}
\def\Dzb     {{\ensuremath{\Dbar{}^0}}\xspace}
\def\Dp      {{\ensuremath{\D^+}}\xspace}
\def\Dm      {{\ensuremath{\D^-}}\xspace}

\def\DpDm    {\ensuremath{\Dp {\kern -0.16em \Dm}}\xspace}

\def\Dstarp  {{\ensuremath{\D^{*+}}}\xspace}

\def\B       {{\ensuremath{\PB}}\xspace}

\def\BorBbar {\kern \thebaroffset\optbar{\kern -\thebaroffset \PB}\xspace}

\def\Bd      {{\ensuremath{\B^0}}\xspace}

\def\BdorBdbar {\kern \thebaroffset\optbar{\kern -\thebaroffset \Bd}\xspace}
\def\Bu      {{\ensuremath{\B^+}}\xspace}

\def\Bp      {{\ensuremath{\Bu}}\xspace}

\def\Bs      {{\ensuremath{\B^0_\squark}}\xspace}

\def\BsorBsbar {\kern \thebaroffset\optbar{\kern -\thebaroffset \Bs}\xspace}

\def\jpsi     {{\ensuremath{{\PJ\mskip -3mu/\mskip -2mu\Ppsi}}}\xspace}

\def\Y#1S{\ensuremath{\PUpsilon{(#1S)}}\xspace}

\def\proton      {{\ensuremath{\Pp}}\xspace}

\def\Lz          {{\ensuremath{\PLambda}}\xspace}

\def\LorLbar     {\kern \thebaroffset\optbar{\kern -\thebaroffset \PLambda}\xspace}

\def\Lc          {{\ensuremath{\Lz^+_\cquark}}\xspace}

\def\Lb           {{\ensuremath{\Lz^0_\bquark}}\xspace}

\def\BF         {{\ensuremath{\mathcal{B}}}\xspace}

\newcommand{\decay}[2]{\mbox{\ensuremath{#1\!\to #2}}\xspace}

\def\to                 {\ensuremath{\rightarrow}\xspace}

\def\AT#1     {\ensuremath{A_{\mathrm{T}}^{#1}}\xspace}           

\def\C#1      {\ensuremath{\mathcal{C}_{#1}}\xspace}                       
\def\Cp#1     {\ensuremath{\mathcal{C}_{#1}^{'}}\xspace}                    
\def\Ceff#1   {\ensuremath{\mathcal{C}_{#1}^{\mathrm{(eff)}}}\xspace}        
\def\Cpeff#1  {\ensuremath{\mathcal{C}_{#1}^{'\mathrm{(eff)}}}\xspace}       
\def\Ope#1    {\ensuremath{\mathcal{O}_{#1}}\xspace}                       
\def\Opep#1   {\ensuremath{\mathcal{O}_{#1}^{'}}\xspace}                    

\newcommand{\aunit}[1]{\ensuremath{\text{\,#1}}}

\newcommand{\tev}{\aunit{Te\kern -0.1em V}\xspace}
\newcommand{\gev}{\aunit{Ge\kern -0.1em V}\xspace}
\newcommand{\mev}{\aunit{Me\kern -0.1em V}\xspace}
\newcommand{\kev}{\aunit{ke\kern -0.1em V}\xspace}
\newcommand{\ev}{\aunit{e\kern -0.1em V}\xspace}

\newcommand{\mevc}{\ensuremath{\aunit{Me\kern -0.1em V\!/}c}\xspace}
\newcommand{\gevc}{\ensuremath{\aunit{Ge\kern -0.1em V\!/}c}\xspace}
\newcommand{\mevcc}{\ensuremath{\aunit{Me\kern -0.1em V\!/}c^2}\xspace}
\newcommand{\gevcc}{\ensuremath{\aunit{Ge\kern -0.1em V\!/}c^2}\xspace}

\def\fb   {\ensuremath{\aunit{fb}}\xspace}
\def\invfb   {\ensuremath{\fb^{-1}}\xspace}

\def\gsim{{~\raise.15em\hbox{$>$}\kern-.85em
          \lower.35em\hbox{$\sim$}~}\xspace}
\def\lsim{{~\raise.15em\hbox{$<$}\kern-.85em
          \lower.35em\hbox{$\sim$}~}\xspace}

\def\pt         {\ensuremath{p_{\mathrm{T}}}\xspace}

\def\evtgen     {\mbox{\textsc{EvtGen}}\xspace}

\def\geant      {\mbox{\textsc{Geant4}}\xspace}

\def\photos     {\mbox{\textsc{Photos}}\xspace}

\def\pythia     {\mbox{\textsc{Pythia}}\xspace}

\def\tell1  {TELL1\xspace}
\def\ukl1   {UKL1\xspace}

\newcommand{\lhcborcid}[1]{\href{https://orcid.org/#1}{\hspace*{0.1em}\raisebox{-0.45ex}{\includegraphics[width=1em]{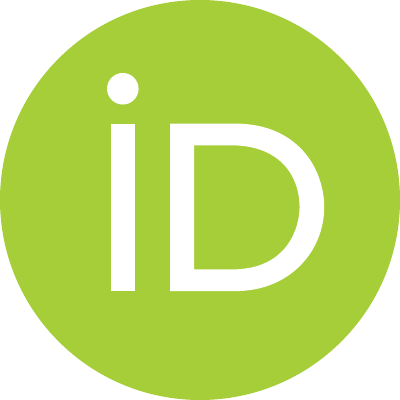}}}}


%% file: local-symbols-def.tex
\newcommand{\Btopimue}{\ensuremath{\decay{\Bu}{\pip\mupm\emp}}\xspace}

\newcommand{\BtoJpsK}{\ensuremath{\decay{\Bu}{\jpsi(\decay{}{\mumu})\Kp}}\xspace}

\newcommand{\BtoJpsKee}{\ensuremath{\decay{\Bu}{\jpsi(\decay{}{\ep \en})\Kp}}\xspace}

%% file: title-LHCb-PAPER.tex

\begin{titlepage}
\pagenumbering{roman}

\vspace*{-1.5cm}
\centerline{\large EUROPEAN ORGANIZATION FOR NUCLEAR RESEARCH (CERN)}
\vspace*{1.5cm}
\noindent
\begin{tabular*}{\linewidth}{lc@{\extracolsep{\fill}}r@{\extracolsep{0pt}}}
\ifthenelse{\boolean{pdflatex}}
{\vspace*{-1.5cm}\mbox{\!\!\!\includegraphics[width=.14\textwidth]{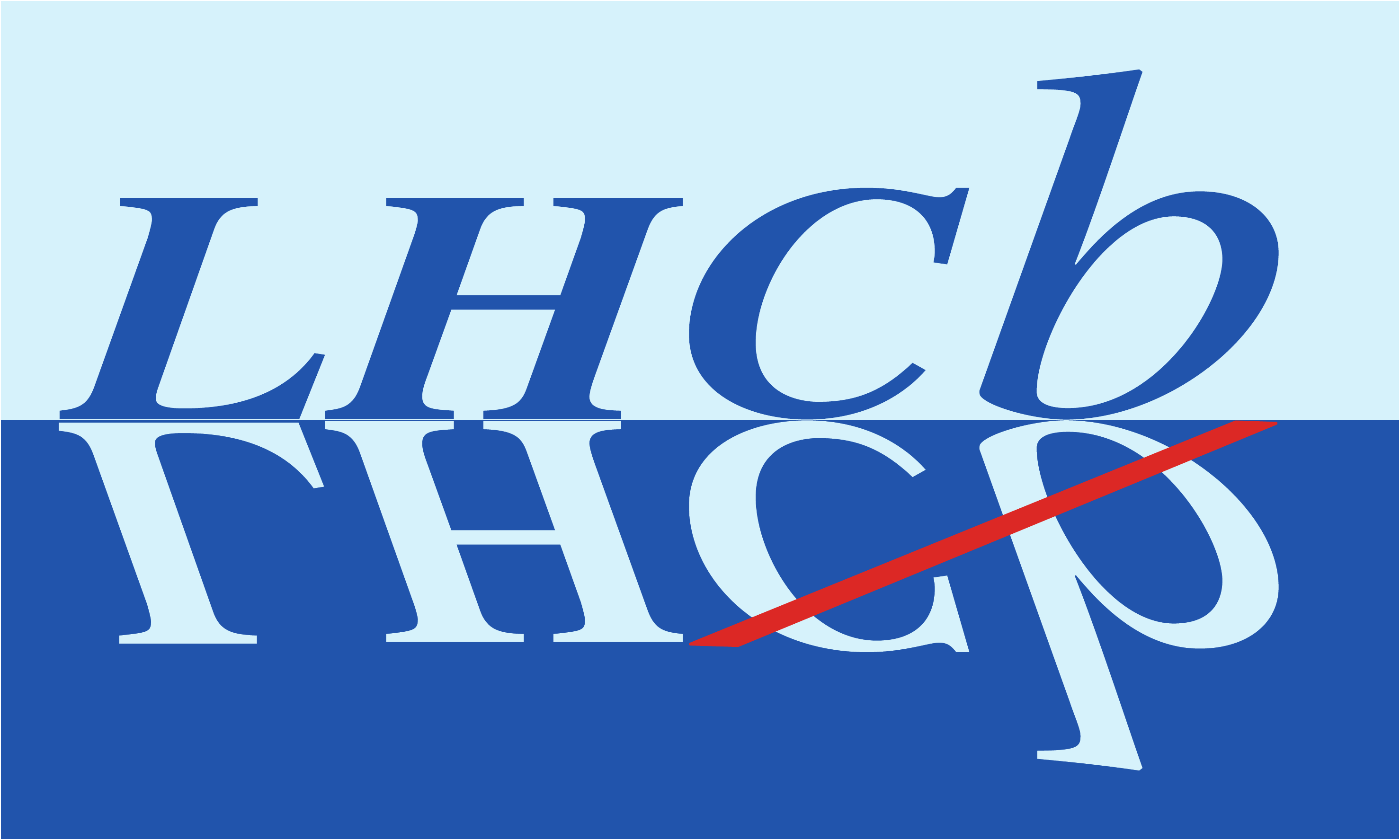}} & &}%
{\vspace*{-1.2cm}\mbox{\!\!\!\includegraphics[width=.12\textwidth]{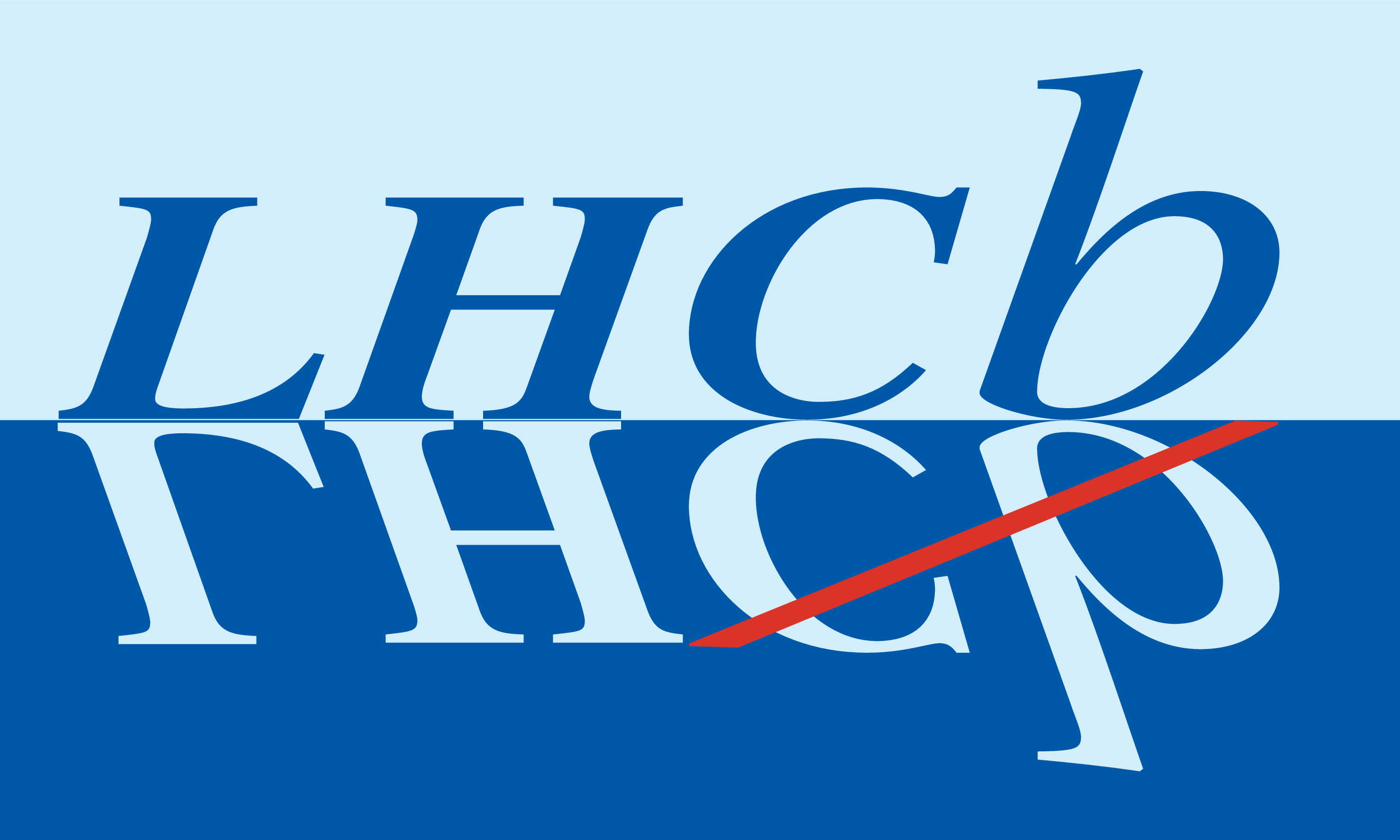}} & &}%
\\
 & & CERN-EP-2026-093 \\  
 & & LHCb-PAPER-2026-013 \\  
 & & April 9, 2026 \\ 
 & & \\
\end{tabular*}

\vspace*{4.0cm}

{\normalfont\bfseries\boldmath\huge
\begin{center}
  \papertitle 
\end{center}
}

\vspace*{2.0cm}

\begin{center}
\paperauthors\footnote{Authors are listed at the end of this Letter.}
\end{center}

\vspace{\fill}

\begin{abstract}
  \noindent
The first search for the lepton-flavour violating decays \Btopimue in proton-proton collisions is presented, using data collected by the \lhcb experiment between 2011 and 2018, corresponding to an integrated luminosity of 9\invfb.  No significant signal is observed and an upper limit on the branching fraction is set at $\mathcal{B}(\Btopimue) < 1.8 \times 10^{-9}$ at the 90\%\ confidence level, two orders of magnitude more restrictive than the current world average. This is the first constraint on lepton-flavour violating \decay{\bquark}{\dquark} quark transitions at the LHC and also sets the most stringent upper limits to date on \decay{\bquark}{\dquark \mupm  \emp} transitions. Limits on left-handed and scalar scenarios beyond the Standard Model are also reported.

\end{abstract}

\vspace*{2.0cm}

\begin{center}
  Submitted to
  Phys.~Rev.~Lett. 
\end{center}

\vspace{\fill}

{\footnotesize 
\centerline{\copyright~\papercopyright. \href{\paperlicenceurl}{\paperlicence}.}}
\vspace*{2mm}

\end{titlepage}


\newpage
\setcounter{page}{2}
\mbox{~}
%
%
%
%

%% file: body.tex
Lepton-flavour violation (LFV) has been firmly established in the neutrino sector through oscillation measurements, but is essentially absent in the charged sector of the Standard Model (SM)  where loop-induced contributions to the branching fraction are suppressed to less than $\mathcal{O}(10^{-50})$~\cite{Calibbi:2017uvl}. Any observation of a charged LFV decay would therefore provide unambiguous evidence for physics beyond the SM (BSM). Rare flavour-changing neutral current processes are particularly sensitive to new interactions. Recent measurements of semileptonic \decay{\bquark}{\squark \ell^+\ell^-} transitions, where $\ell$ denotes an electron or muon, have revealed intriguing tensions with SM predictions~\cite{Capdevila:2023yhq}, motivating searches for LFV decays, such as \Btopimue, that could arise from related new dynamics. 
Although these modes are unobservable in the SM, several BSM scenarios including models with leptoquarks~\cite{Duraisamy:2016gsd}, extended gauge sectors~\cite{Crivellin:2015era}, or non-minimal Higgs structures~\cite{Hou:2022mwr} could enhance the branching fraction to $\mathcal{O}(10^{-10})$.
The \lhcb collaboration has previously searched for charged LFV in semileptonic \decay{\bquark}{\squark \ellpm \ell^{\prime \mp}} transitions, setting stringent upper limits~\cite{LHCb-PAPER-2019-022, LHCb-PAPER-2019-043, LHCb-PAPER-2022-008, LHCb-PAPER-2022-021, LHCb-PAPER-2024-006, LHCb-PAPER-2025-005}. The less-explored \decay{\bquark}{\dquark} quark transition offers complementary sensitivity to possible BSM contributions.
The most stringent upper limits to date on $\BF(\Btopimue)$ (at 90\% CL) are set by the \cleo  and the \babar collaborations at  
$1.6 \times 10^{-6}$ and $1.7 \times 10^{-7}$, respectively~\cite{CLEO:2002cel, BaBar:2007xeb}.

This Letter presents the first search for the rare decays \Btopimue at the LHC, using proton-proton ($pp$) collision data collected by the \lhcb experiment between 2011 and 2018 at centre-of-mass energies of 7, 8 and 13\tev, corresponding to an integrated luminosity of approximately 9\invfb. The inclusion of charge-conjugate processes is implied throughout this Letter. The branching fraction of the signal decay is determined relative to the well-measured  \BtoJpsK decay, which serves as the normalisation mode due to its similar kinematic topology and experimental signature.


The \lhcb detector~\cite{LHCb-DP-2008-001,LHCb-DP-2014-002} is a single-arm forward spectrometer covering the pseudorapidity range $2<\eta<5$, designed for the study of particles containing \bquark or \cquark quarks. The detector used to collect the data analysed in this Letter includes a high-precision tracking system consisting of a silicon-strip vertex detector surrounding the $pp$ interaction region~\cite{LHCb-DP-2014-001}, a large-area silicon-strip detector located upstream of a dipole magnet with a bending power of about $4{\mathrm{\,T\,m}}$, and three stations of silicon-strip detectors and straw drift tubes~\cite{LHCb-DP-2013-003,LHCb-DP-2017-001} placed downstream of the magnet. The tracking system provides the measurements of the track momentum and impact parameter (IP) and is used to reconstruct primary vertices (PVs). 
Different types of charged hadrons are distinguished using particle identification (PID) information from two ring-imaging Cherenkov detectors~\cite{LHCb-DP-2012-003}.
Photons, electrons and hadrons are identified by a calorimeter system consisting of scintillating-pad and preshower detectors, an electromagnetic 
and a hadronic calorimeter.
Muons are identified by a system composed of alternating layers of  iron and multiwire proportional chambers~\cite{LHCb-DP-2012-002}.

The online event selection is performed by a trigger~\cite{LHCb-DP-2012-004}, which consists of a hardware stage, based on information from the muon system, followed by a two-level software stage, which performs a full event reconstruction. At the hardware trigger stage, events are required to contain a muon with high transverse momentum (\pt).
The first-level software trigger requires one or two tracks that 
have large \pt\ and are significantly displaced from every PV. 
The second-level software trigger requires a two-, three- or four-track secondary vertex with a significant displacement from any primary $pp$ interaction vertex.  A multivariate algorithm~\cite{LHCb-PROC-2015-018} is used for the identification of secondary vertices that are consistent with the decay of a \bquark hadron. Triggered data further undergo a centralised, offline processing step to make them suitable for physics analysis across the entire \lhcb physics programme~\cite{Stripping}.

Simulated data are used to study the properties of the signal, normalisation and background channels. 
The generation of $pp$ collisions is performed using \pythia~\cite{Sjostrand:2007gs}, with a specific \lhcb configuration~\cite{LHCb-PROC-2010-056}. Decays of unstable particles are described by \evtgen~\cite{Lange:2001uf}, in which final-state radiation is generated using \photos~\cite{davidson2015photos}.
Interactions of the generated particles with the detector material and the response of the detector are implemented using the \geant toolkit~\cite{Allison:2006ve, *Agostinelli:2002hh}, as described in Ref.~\cite{LHCb-PROC-2011-006}. A uniform phase-space decay model is used to generate the simulated signal samples.
The simulated data are corrected using the \BtoJpsK decay channel to account for known imperfections in modelling of the \Bp production kinematics and detector occupancy, as well as track reconstruction and trigger efficiencies. 
Data calibration samples consisting of high purity \decay{\Bp}{\jpsi (\decay{}{\ep\en})X}, \decay{\Dstarp}{\Dz (\decay{}{\Km \pip})\pip} and \decay{B}{\jpsi (\decay{}{\mumu})X} decays are used to calculate PID efficiencies for electrons, muons and hadrons, respectively,  as functions of track momentum and pseudorapidity~\cite{LHCb-PUB-2016-021}. Here, $X$ represents one or more additional hadrons.


The \Btopimue decay candidates satisfying the trigger requirements are reconstructed by combining three tracks that originate from a common vertex with good fit quality that is well separated from any PV. Further selections are applied to ensure the tracks are of a good quality and contain hits in the relevant subdetectors that provide particle identification. The mass of the reconstructed \Bp candidate, $m( \pip \mupm \emp )$, must lie in the range [4500, 6000]\mevcc. To avoid potential bias, the signal region [4985, 5385]\mevcc was not examined until the full procedure had been finalised.
Photons are associated with reconstructed electron trajectories to take into account potential bremsstrahlung energy losses incurred while passing through the \lhcb detector. 
As bremsstrahlung emission affects the mass distribution and kinematics of signal candidates, data are classified according to the presence or absence of bremsstrahlung photons associated with the electron candidate. These are further divided by  data-taking periods 2011--2012 and 2015--2018, giving four analysis categories that are treated independently during selection. 

To suppress combinatorial background arising from unrelated tracks that are incorrectly combined into a common vertex, a boosted decision tree~(BDT) classifier~\cite{Breiman,AdaBoost} is used. This is trained using simulated signal events that satisfy the preceding selection requirements, together with data from the upper mass sideband, $m( \pip \mupm \emp ) \in [5385, 6000]\mevcc$, which serves as a proxy for the combinatorial background. This BDT exploits observables describing the \Bp candidate kinematics, as well as those related to its decay vertex quality, its IP with respect to the associated PV and the angle between its flight direction and momentum vector. The PV that is most consistent with the flight direction of the \Bp candidate is taken as the associated PV.
    Variables related to the final-state particles are also used and  include  the \pt of the pion, the IP of each track with respect to the PV and the angular separation,  $\Delta R \equiv \sqrt{(\Delta \eta)^2 + (\Delta \phi)^2}$, between the pion and each lepton, where $\Delta \eta$ and $\Delta \phi$ are the pseudorapidity and azimuthal angle differences between the two tracks.
The BDT also uses the response of a dedicated track-isolation algorithm, introduced in the search for \decay{\Bs}{\mup \mun} decays~\cite{LHCb-PAPER-2017-001}. The BDT training employs a $k$-fold cross-validation scheme with five folds~\cite{Bevan:2017stx}, which allows the full sample to be used for training while avoiding biases in the classifier output. The classifier performance is  optimised via a Bayesian search of its hyperparameter space using the \texttt{Optuna} package~\cite{optuna_2019},
with a figure of merit designed to  maximise the discrimination power of the classifier while minimising overtraining.
The selection on the BDT response is determined independently for each analysis category by retaining 75\% signal efficiency, as evaluated using the simulated signal samples.

To remove residual combinatorial and partially reconstructed backgrounds present in the lower mass sideband, a second BDT is trained and applied after the first. This BDT is trained on the signal simulation and data from the lower mass sideband corresponding to $m( \pip \mupm \emp ) \in [4500, 4985]\mevcc$.
The training shares the same $k$-folding approach and hyperparameter optimisation as the first BDT, using a subset of the discriminating variables as well as the ratio of the  electron and the combined pion and muon momenta components transverse to the \Bp direction of flight.
The selection on the response from the second BDT is optimised simultaneously with the requirements on the PID variables for each of the final-state tracks, using a figure of merit which is well suited for search analyses~\cite{Punzi:2003bu}, $\varepsilon/(\sqrt{n} + 3/2)$, where $\varepsilon$ is the signal efficiency estimated from simulation and $n$ is the estimated background yield in the signal region. 
This optimisation is performed separately in each analysis category through a grid search of the parameters of interest
and results in a selection efficiency that varies between 20\% and 40\%.
At most 1\% of selected events have multiple candidates; in such events, one candidate is chosen randomly. The selection of \BtoJpsK decays follows that of the signal mode where possible. The optimal PID requirement defined for the signal mode muon candidate is applied to both leptons, while the hadron PID requirement is inverted. The second BDT is not applied due to the absence of a final-state electron.

Dedicated simulated samples of exclusive \bquark-hadron decays are used to assess possible background contamination in the signal region after the full selection. These samples are corrected using the same procedure as for the signal and normalisation modes. Samples consisting of fully reconstructed decays such as \decay{\Bp}{\jpsi (\decay{}{\ellp \ellm}) \Kp}, \decay{\Bp}{\pip \ellp \ellm} and \decay{\Bp}{\pip \pip \pim} and partially reconstructed decays such as \decay{\Bp}{\jpsi (\decay{}{\ep \en}) \Kstarp (\decay{}{\Kp\piz})}, \decay{\Lb}{\Lc (\decay{}{\Lz (\decay{}{\proton \pim}) \ep \neue }) \mun \neumb} and \decay{\Bp}{\Dzb \ellp \neul}, where \decay{\Dzb}{\pip \ellm \neulb} or \decay{\Dzb}{\Kp \pim}, 
are considered. In all cases, some final-state particles are either not reconstructed or are  misidentified. Contributions from all other possible lepton-flavour violating decays are assumed to be zero. The expected yield for each exclusive background decay is calculated relative to the number of \BtoJpsK decays observed in data after the full selection. Although small contributions are observed in the lower mass sideband, no background mode leads to a peaking structure close to the expected signal mass. 

To account for potential residual mismodelling arising from imperfect simulation or limited experimental knowledge of specific decay modes, a data-driven method is employed to estimate contributions from all possible partially reconstructed and misidentified backgrounds. The method, originally developed for electrons in Ref.~\cite{LHCb-PAPER-2022-046}, involves inverting the PID requirement on all possible combinations of tracks, thereby defining seven background-enriched regions. Each of these regions is weighted to parameterise the misidentification efficiencies of kaons, pions and electrons. The weights are obtained from the calibration samples mentioned above, subjected to the same selection criteria as the signal sample. All  weighted background-enriched regions are combined  to obtain an estimate of the residual background mass distribution in the signal region. The invariant-mass shape of these backgrounds is modelled using a kernel density estimator~\cite{CRANMER2001198} and  included as a component in the upper limit setting procedure.

The signal branching fraction $\BF(\Btopimue)$ is determined relative to the normalisation mode using
\begin{align}
\BF(\Btopimue) &= N(\Btopimue) \cdot \alpha\,, \nonumber\\
                \alpha &\equiv \frac{\BF(\BtoJpsK)}{\varepsilon \left( \Btopimue \right)} \frac{\varepsilon\left( \BtoJpsK \right)}{N\left( \BtoJpsK \right)}\,, 
\label{eqn:branching_fraction}
\end{align}
where $\varepsilon(Y)$ and $N(Y)$ represent the efficiency and the observed yield, respectively, for either signal or normalisation mode $Y$.
The branching fraction of the normalisation mode is taken as 
    $(6.08 \pm 0.12)\times 10^{-5}$~\cite{PDG2024}. The yield of the normalisation channel is obtained from an unbinned extended maximum-likelihood fit to the invariant mass, $m(\Kp \mup \mun)$, of the selected \BtoJpsK candidates, where the invariant mass of the dimuon pair is constrained to that of the known \jpsi mass and the sum of all final-state particle momenta is constrained to point back to the associated PV. The mass distribution is modelled using a double-sided Crystal Ball function~\cite{Skwarnicki:1986xj} with power-law tails on both sides of the peak, resulting in a yield of $(1.3609 \pm 0.0012)\times 10 ^ {6}$ candidates. The contribution from combinatorial background is demonstrated to be negligible. The value of the normalisation constant, $\alpha$, is calculated as $(3.58 \pm 0.07)\times 10^{-10}$, where the uncertainty is statistical only.

\begin{figure}[tbp]
    \centering
    \includegraphics[width=0.6\textwidth]{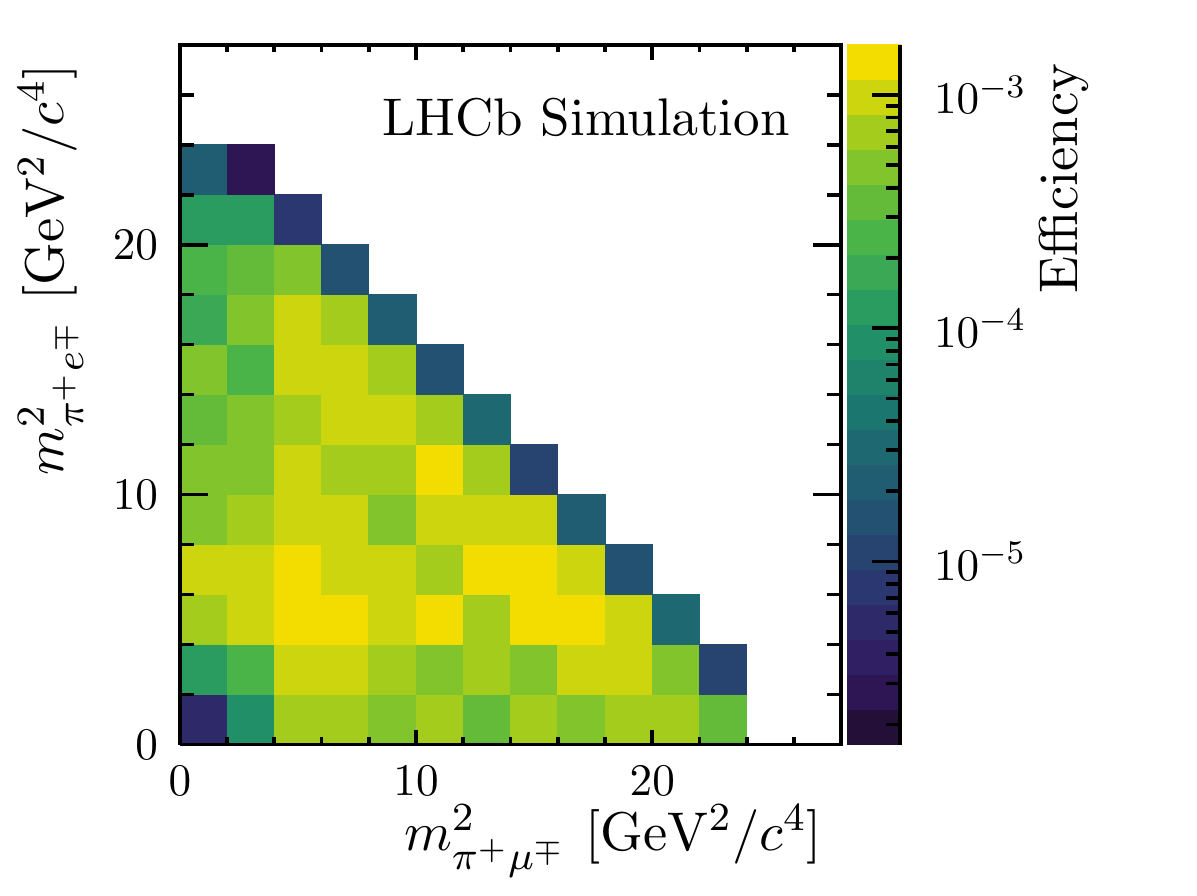}
    \caption{Selection efficiency of \Btopimue decays as a function of the squared invariant masses, $m^2_{\pi^+ \mu^\pm}$ and $m^2_{\pi^+ e^\mp}$. The efficiency varies across the Dalitz plane due to the applied selection requirements.}
    \label{fig:dalitz-eff}
\end{figure}

The efficiencies are evaluated taking into account all selection requirements. As the efficiency is not uniform across the phase space of the signal decay, it is calculated in intervals of the squared invariant masses of the final-state particles, $m^2_{\pi^+ \mu^\pm}$ and $m^2_{\pi^+ e^\mp}$, as shown in Fig.~\ref{fig:dalitz-eff} and available in  Ref.~\cite{HEPData_entry}. 
This can be used to reinterpret the result for alternative BSM models that can have different \Btopimue decay kinematics.

Potential sources of systematic uncertainties, primarily from efficiency corrections and the modelling of background components, are summarised in Table~\ref{tab:syst_summary}.
Most selection-related effects cancel in the ratio of Eq.~\ref{eqn:branching_fraction} and therefore only small effects are expected.
Differences between data and simulation in both the first and second BDT selections are quantified using background-subtracted \BtoJpsK and \BtoJpsKee decays, respectively, resulting in uncertainties of up to 2.0\% depending on the analysis category.
The finite size of the simulated samples results in an uncertainty of about 1.3\%. 
Uncertainty on the normalisation mode branching fraction contributes an additional 1.9\%.
The choice of signal and background parameterisations also contributes  systematic uncertainties  of  1.8\% and 6.6\%, respectively.
Systematic effects arising from the correction of known imperfections in the simulation of the \Bp production kinematics and PID, tracking and hardware trigger efficiencies are summed in quadrature.
Efficiencies for PID are re-evaluated using both coarser and finer binning schemes of the calibration samples, resulting in changes of 0.3--3.5\%, depending on the analysis category. Tracking efficiencies are evaluated using alternative parameterisations of the calibration sample, producing variations of 1.0--1.5\%. The hardware trigger efficiency correction is calculated using various two-dimensional intervals in \pt and either $\eta$ or event multiplicity, resulting in effects below 0.2\%. Variations in the weighting model used to account for production kinematics lead to changes of up to 1.5\%.
The parameterisation of the normalisation channel mass distribution has a negligible impact.

\begin{table}[tb]
	\centering
	\caption{Sources of systematic uncertainties relative to the measured branching fraction, computed as weighted averages across analysis categories.}
	\begin{tabular}{lc}
		Source &  Uncertainty (\%) \\
		\hline
                Data--simulation differences & {$1.3$}  \\
                Simulation sample size  & {$1.3$}  \\
                External branching fraction & {$1.9$}  \\
                Signal model & {$1.8$}  \\
                Background model & {$6.6$}  \\
                Simulation corrections & {$2.5$}  \\
                
    \hline
    Total & {$7.7$} \\
	\end{tabular}
	\label{tab:syst_summary}
\end{table}

\begin{figure}[tbp]
        \centering
        \includegraphics[width=0.48\textwidth]{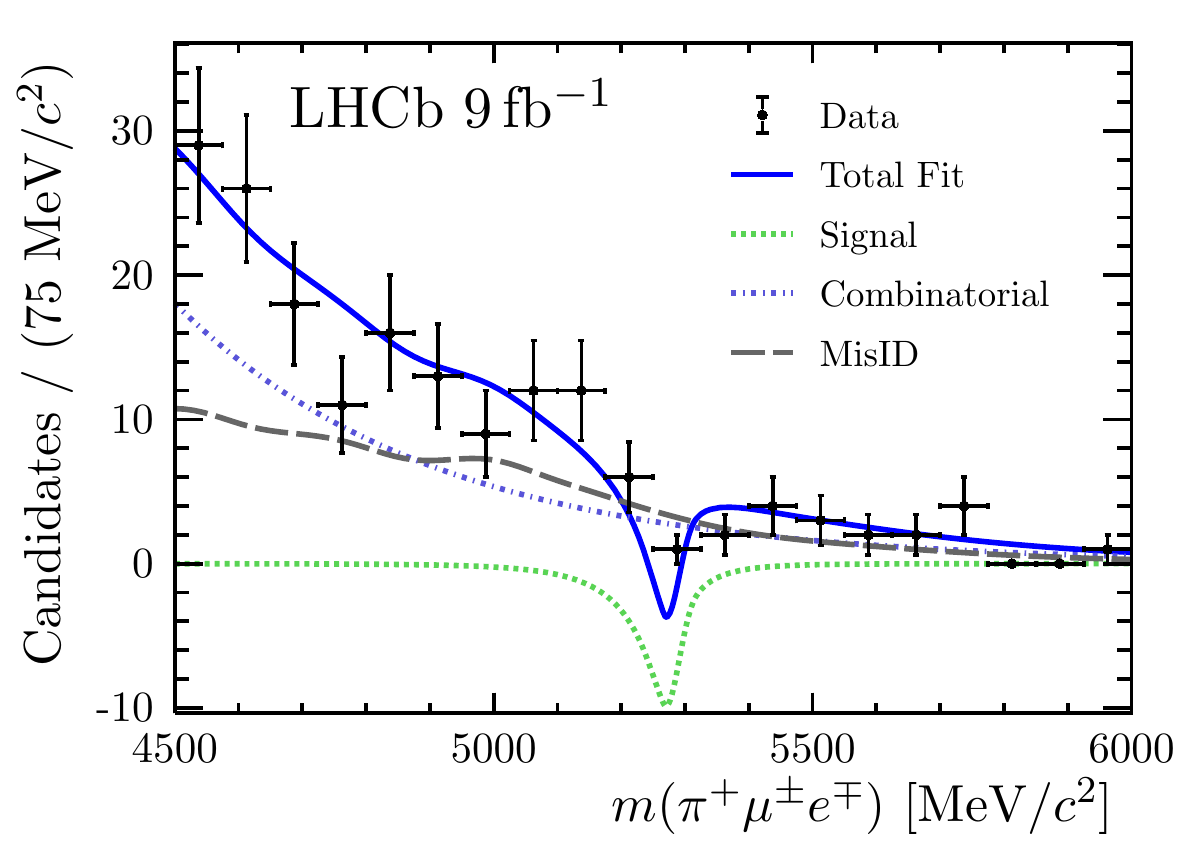}
        \includegraphics[width=0.48\textwidth]{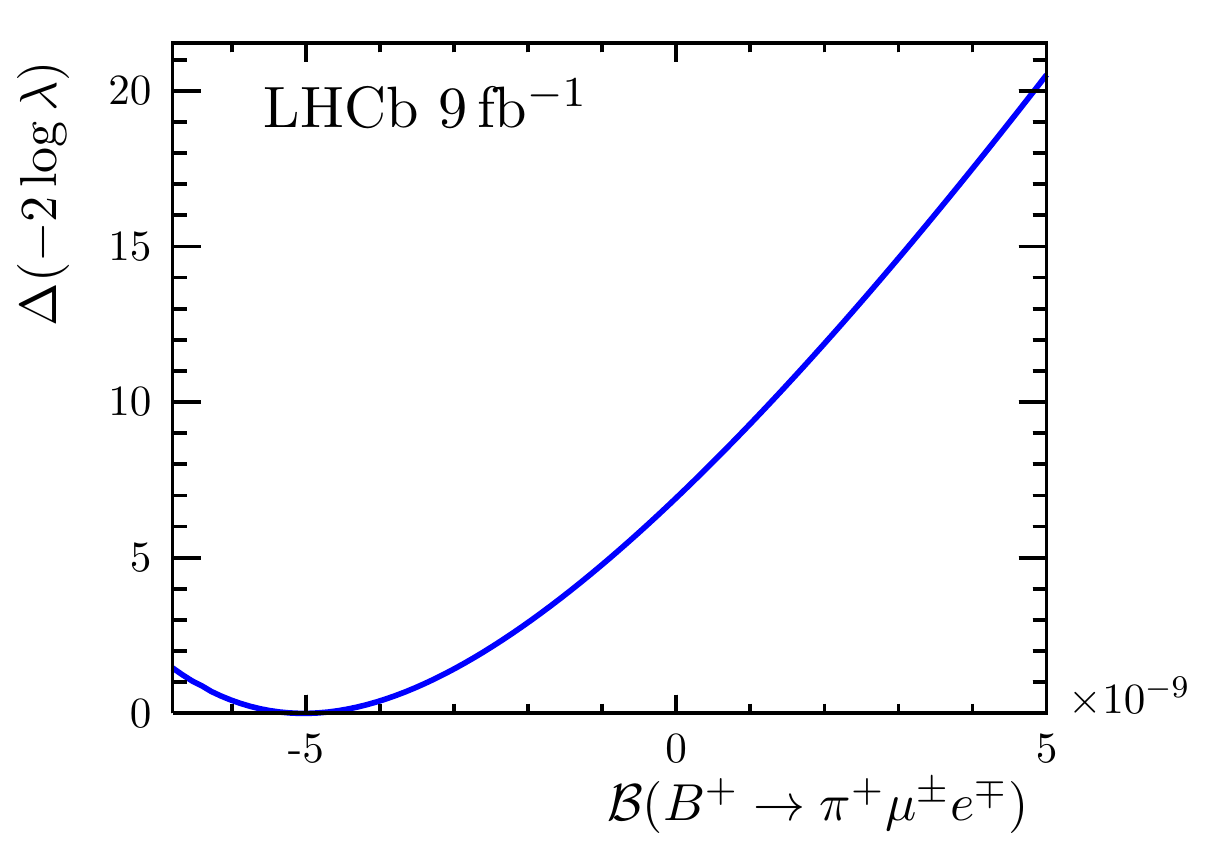}
    \caption{(Left) Invariant-mass distribution of \Btopimue candidates 
    with the signal-plus-background fit function also shown. (Right) Profile likelihood-ratio scan for \Btopimue branching fraction with all nuisance parameters profiled for the baseline signal model fit to data, available at Ref.~\cite{HEPData_entry}.}
    \label{fig:unblinded-mass-dist}
\end{figure}

\begin{figure}[tbp]
    \centering
    \includegraphics[width=0.6\textwidth]{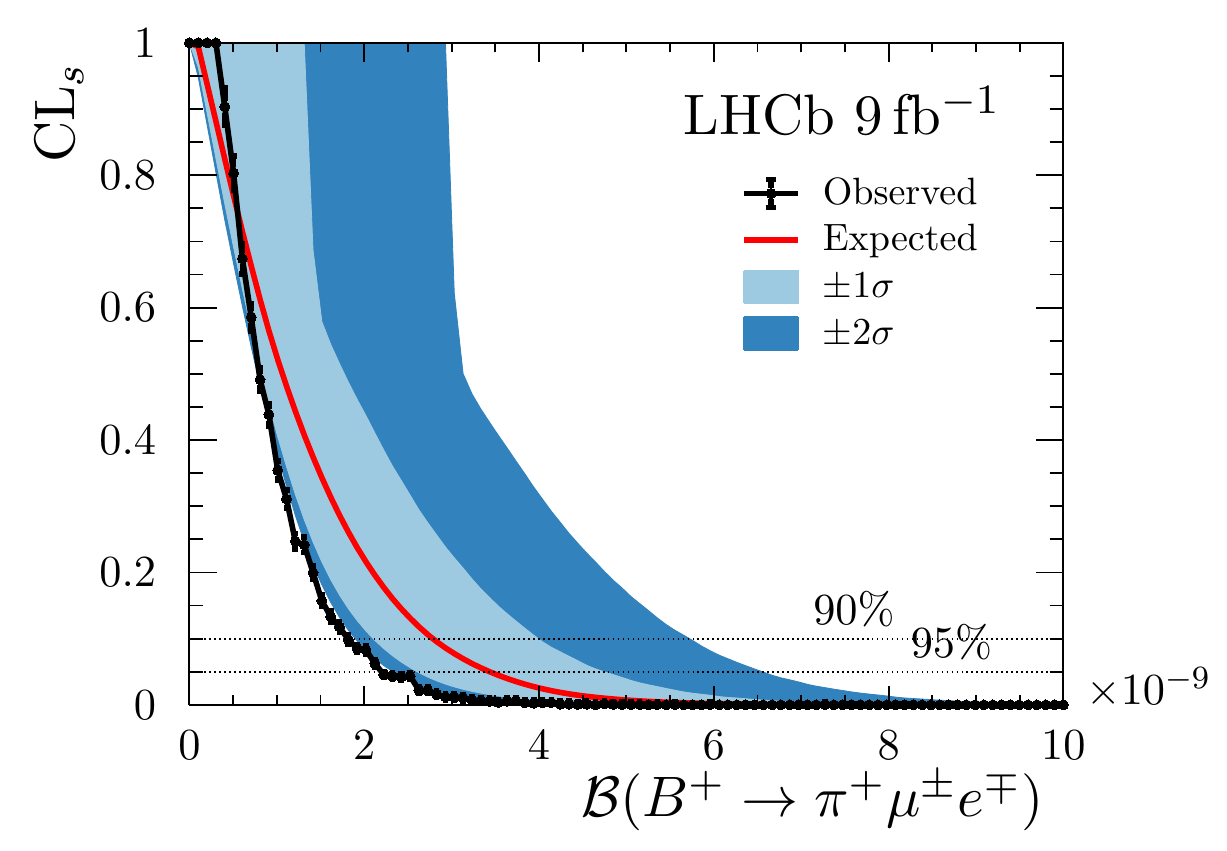}
    \caption{Upper limit on the branching fraction of the \Btopimue decay  obtained on the combined data sets. The red solid line (black data points) corresponds to the expected (observed) $\text{CL}_s$ $p$-value at each value of the branching fraction, and the light blue (dark blue) band indicates the $1\sigma$ ($2\sigma$) uncertainties.}
    \label{fig:cls-scan}
\end{figure}

Figure~\ref{fig:unblinded-mass-dist}  presents the candidate mass distribution summed over all analysis categories, with $36$ ($41 \pm 3$) candidates observed (expected) in the signal region, $m(\pip \mupm \emp) \in [4985, 5385]\mevcc$.
The expected background yield is unchanged by the inclusion of data in the signal region.
Since the data are consistent with the background-only hypothesis, 
an upper limit on $\BF(\Btopimue)$ is set using the $\text{CL}_s$ method~\cite{CLs} with a one-sided test statistic~\cite{GammaCombo}. The likelihood is derived from a simultaneous fit to the invariant-mass distribution in each analysis category with the branching fraction as a shared parameter. The fit comprises three components: a double-sided Crystal Ball function to model the signal, with shape parameters  fixed from simulated signal samples; partially reconstructed and misidentified decays obtained from the aforementioned data-driven method; an exponential function describing the combinatorial background, with its shape parameters determined from a fit to a combinatorial proxy sample. This proxy consists of candidates formed from a muon and electron of the same charge combined with a charged pion. The normalisation constants,  slope of the exponential component and yield of misidentified decays are treated as nuisance parameters, constrained by Gaussian terms that account for their statistical and systematic uncertainties, where applicable.
Pseudoexperiments containing background-only and signal-plus-background mass distributions are generated for varying values of the branching fraction and are fitted to construct distributions of the test statistic, $q = -2\ln\mathcal{\lambda}$, where $\mathcal{\lambda}$ is the profile likelihood ratio of the signal-plus-background hypothesis to the background-only hypothesis with all nuisance parameters profiled. The corresponding profile likelihood-ratio scan is shown in Fig.~\ref{fig:unblinded-mass-dist}. The $\text{CL}_s$ $p$-value is defined as the fraction of pseudoexperiments yielding a value of $q$ larger than that obtained from the fit to data.
The resulting scan is shown in Fig.~\ref{fig:cls-scan}.  The observed upper limits are determined to be $\BF(\Btopimue) < 1.8\,(2.2)\times 10^{-9}$ at the 90\% (95\%) confidence level, while the expected limits are $2.7\,(3.4)\times 10^{-9}$ at the 90\% (95\%) confidence level, assuming a uniform phase-space distribution for the signal decays.
The measured branching fraction is $(-5.0^{+1.7}_{-1.5}) \times 10^{-9}$, consistent with absence of signal. 
However, alternative BSM scenarios can result in different decay kinematics and differential decay rates~\cite{Becirevic:2024vwy}. This is illustrated in Figs.~\ref{fig:left_handed_and_scalar_dll} and~\ref{fig:left_handed_and_scalar_cls_scan} of the End Matter for two distinct BSM scenarios: a left-handed model with ${\cal C}^{\mu e}_{9} = -{\cal C}^{\mu e}_{10} \neq 0$, and a scalar model with ${\cal C}^{\mu e}_{S} \neq 0$, where ${\cal C}^{\mu e}_i$ denotes the lepton-flavour violating Wilson coefficients. The corresponding upper limits for these two models at the 90\%\,(95\%) confidence level are found to be $\mathcal{B}(\Btopimue) < 1.8\,(2.2) \times 10^{-9}$ and $\mathcal{B}(\Btopimue) < 1.7\,(2.1) \times 10^{-9}$, respectively.


In summary, the first search for the lepton-flavour violating decays \Btopimue in $pp$ collisions is performed. The data used were collected between 2011 and 2018 at centre-of-mass energies of $7$, $8$ and $13\tev$, corresponding to an integrated luminosity of approximately 9\invfb. 
No excess is observed relative to the background-only hypothesis and a resulting upper limit on the branching fraction is set 
assuming a uniform distribution of signal decays within the phase space accessible to the final-state particles. This result is  two orders of magnitude more stringent than the previous limit on this decay and provides the first constraint on lepton-flavour violating \decay{\bquark}{\dquark} quark transitions at the LHC. These are the most restrictive upper limits to date on \decay{\bquark}{\dquark \mupm  \emp} transitions. New limits on left-handed and scalar BSM scenarios are also reported.


%% file: acknowledgements.tex
\section*{Acknowledgements}
%
%
\noindent We express our gratitude to our colleagues in the CERN
accelerator departments for the excellent performance of the LHC. We
thank the technical and administrative staff at the LHCb
institutes.
We acknowledge support from CERN and from the national agencies:
ARC (Australia);
CAPES, CNPq, FAPERJ and FINEP (Brazil); 
MOST and NSFC (China); 
CNRS/IN2P3 and CEA (France);  
BMFTR, DFG and MPG (Germany);
INFN (Italy); 
NWO (Netherlands); 
MNiSW and NCN (Poland); 
MEC/IFA (Romania); 
MICIU and AEI (Spain);
SNSF and SER (Switzerland); 
NASU (Ukraine); 
STFC (United Kingdom); 
DOE NP and NSF (USA).
We acknowledge the computing resources that are provided by ARDC (Australia), 
CBPF (Brazil),
CERN, 
IHEP and LZU (China),
IN2P3 (France), 
KIT and DESY (Germany), 
INFN (Italy), 
SURF (Netherlands),
Polish WLCG (Poland),
IFIN-HH (Romania), 
PIC (Spain), CSCS (Switzerland), 
GridPP (United Kingdom),
and NSF (USA).  
We are indebted to the communities behind the multiple open-source
software packages on which we depend.
Individual groups or members have received support from
RTP (Australia), 
Key Research Program of Frontier Sciences of CAS, CAS PIFI, CAS CCEPP (China); 
Minciencias (Colombia);
EPLANET, Marie Sk\l{}odowska-Curie Actions, ERC and NextGenerationEU (European Union);
A*MIDEX, ANR, IPhU and Labex P2IO, and R\'{e}gion Auvergne-Rh\^{o}ne-Alpes (France);
Alexander-von-Humboldt Foundation (Germany);
ICSC (Italy); 
Severo Ochoa and Mar\'ia de Maeztu Units of Excellence, GVA, XuntaGal, GENCAT, InTalent-Inditex and Prog.~Atracci\'on Talento CM (Spain);
the Leverhulme Trust, the Royal Society and UKRI (United Kingdom).

%% file: End_Matter.tex
\clearpage
\section*{End Matter}

\section*{Branching fraction upper limit in BSM scenarios}

\begin{figure}[h!]
    \centering
    \includegraphics[width=0.48\textwidth]{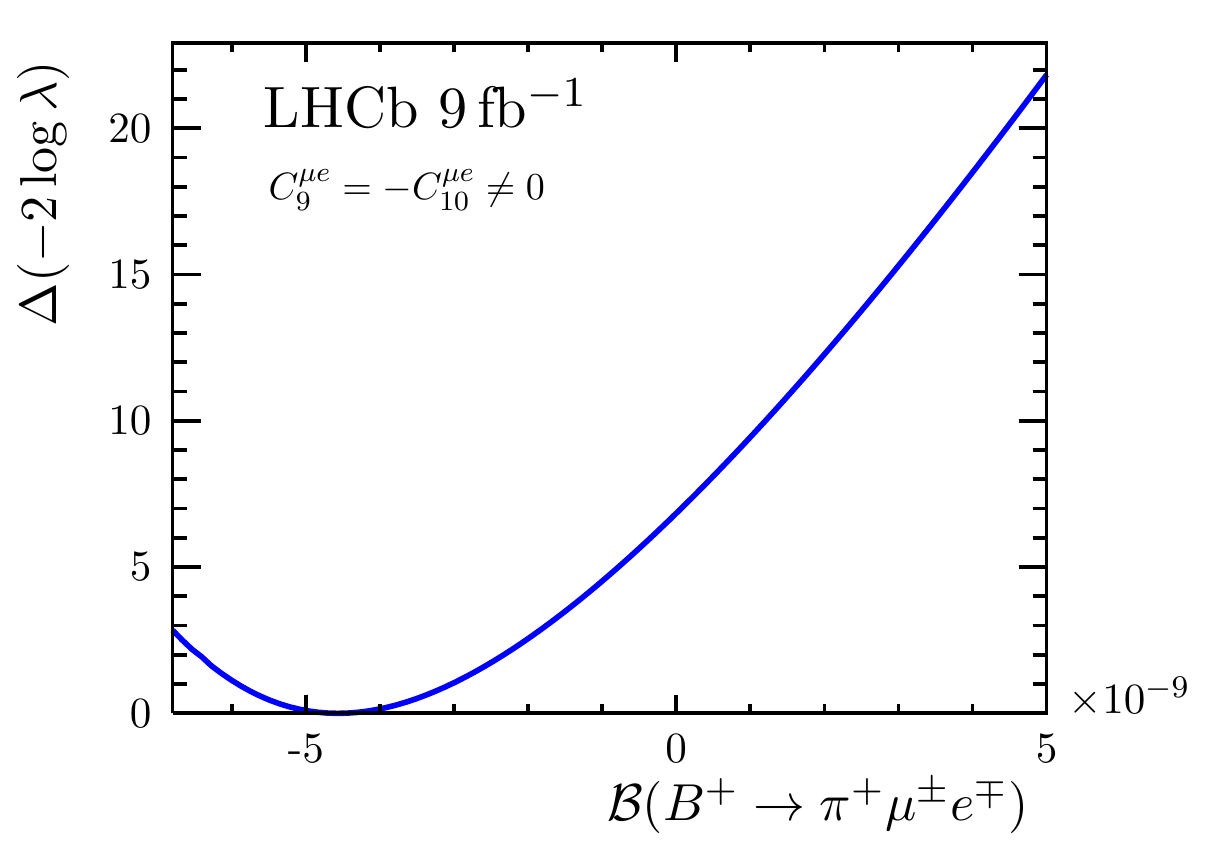}
    \includegraphics[width=0.48\textwidth]{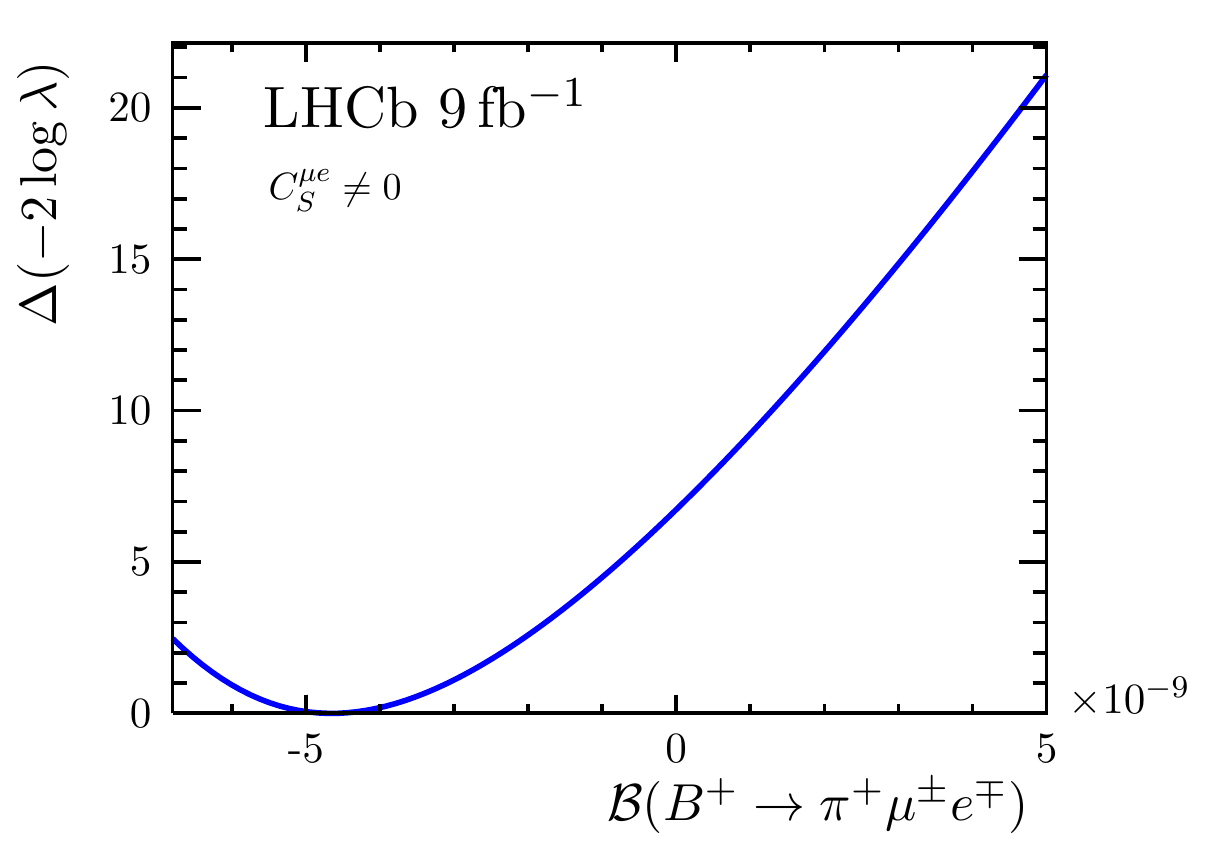}
    \caption{Profile likelihood-ratio scan for \Btopimue branching fraction with all nuisance parameters profiled for (left) left-handed and (right) scalar signal model fit to data, available at Ref.~\cite{HEPData_entry}.}
    \label{fig:left_handed_and_scalar_dll}
\end{figure}

\begin{figure}[h!]
    \centering
    \includegraphics[width=0.48\textwidth]{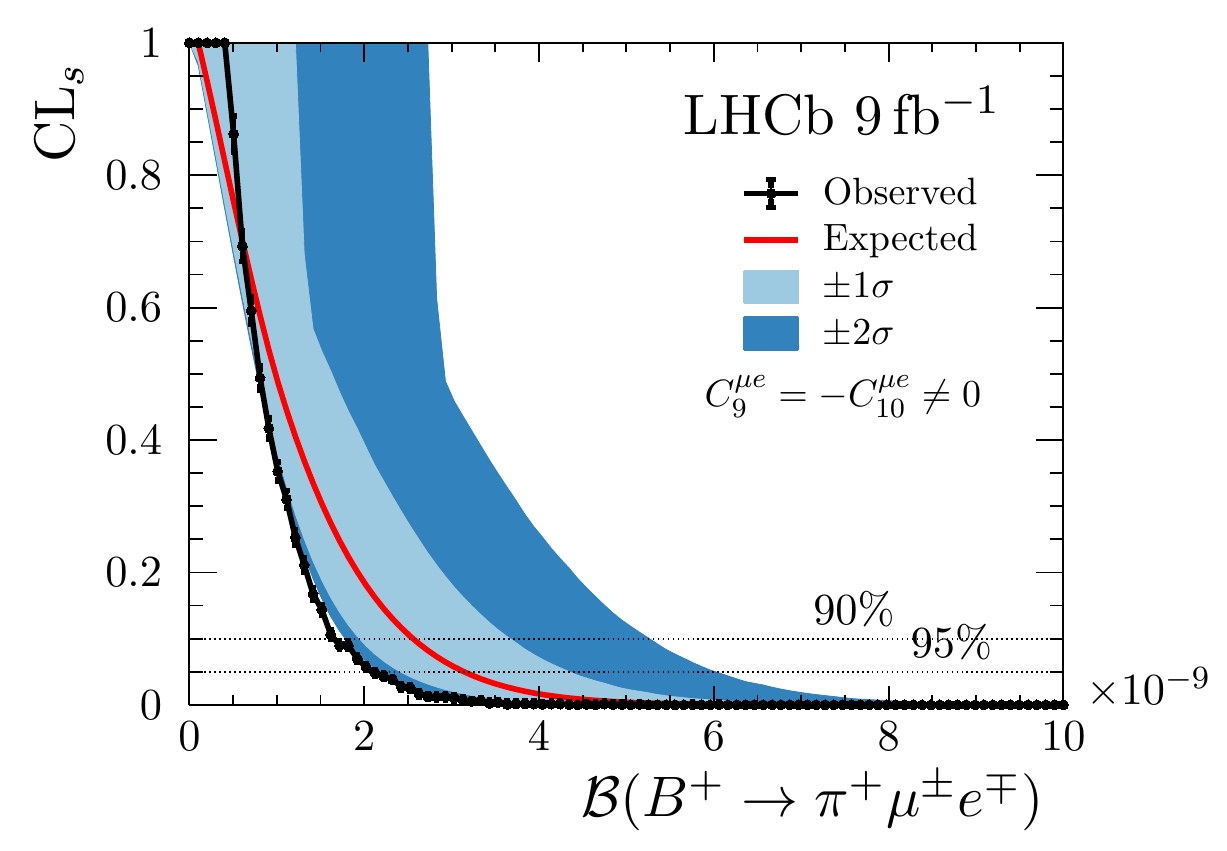}
    \includegraphics[width=0.48\textwidth]{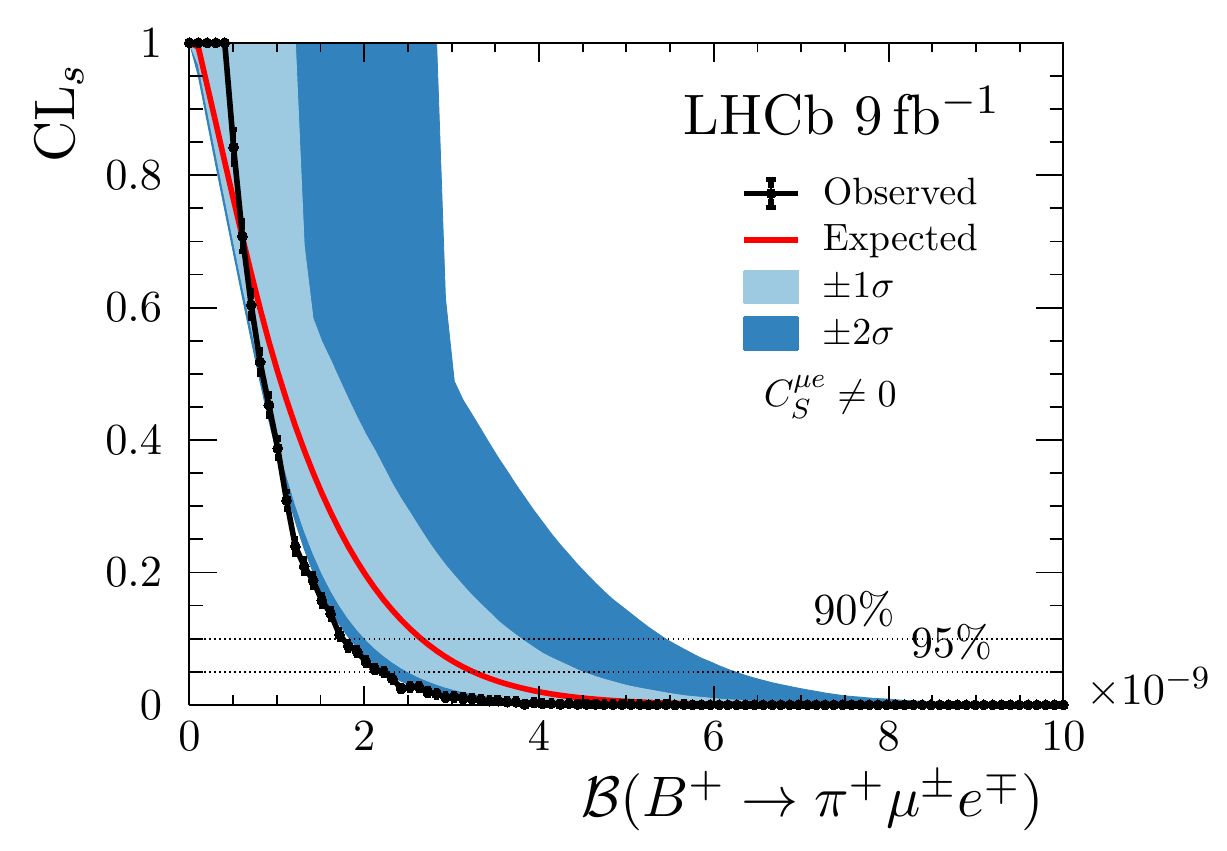}
    \caption{Upper limits on the branching fraction of the \Btopimue decay with the alternative (left) left-handed and (right) scalar BSM scenarios. The red solid line (black data points) corresponds to the expected (observed)  $\text{CL}_s$ $p$-value at each value of the branching fraction, and the light blue (dark blue) band contains the $1\sigma$ ($2\sigma$) uncertainties.}
    \label{fig:left_handed_and_scalar_cls_scan}
\end{figure}

The limits presented in this section are obtained by reinterpreting the data under benchmark  BSM scenarios that modify the decay kinematics of the \Btopimue decays. The alternative models chosen are a left-handed model with ${\cal C}^{\mu e}_{9} = -{\cal C}^{\mu e}_{10} \neq 0$, and a scalar model with ${\cal C}^{\mu e}_{S} \neq 0$, where ${\cal C}^{\mu e}_i$ denotes the lepton-flavour violating Wilson coefficients. In contrast to the uniform phase-space model used for the baseline result, these scenarios predict a nonuniform population of the Dalitz plane, leading to variations in the signal efficiency due to detector acceptance and selection requirements. Signal efficiencies are therefore recomputed using weights applied to the samples generated with the phase-space model for the corresponding BSM scenario, and the complete $\text{CL}_s$ limit-setting procedure is repeated with the modified signal shapes. The resulting profile-likelihood scans and corresponding upper limits are shown in Figs.~\ref{fig:left_handed_and_scalar_dll} and~\ref{fig:left_handed_and_scalar_cls_scan}, respectively.  

%% file: Authorship_LHCb-PAPER-2026-013.tex
\centerline
{\large\bf LHCb collaboration}
\begin
{flushleft}
\small
R.~Aaij$^{38}$\lhcborcid{0000-0003-0533-1952},
M. ~Abdelfatah$^{69}$,
A.S.W.~Abdelmotteleb$^{57}$\lhcborcid{0000-0001-7905-0542},
C.~Abellan~Beteta$^{51}$\lhcborcid{0009-0009-0869-6798},
F.~Abudin\'en$^{59}$\lhcborcid{0000-0002-6737-3528},
T.~Ackernley$^{61}$\lhcborcid{0000-0002-5951-3498},
A. A. ~Adefisoye$^{69}$\lhcborcid{0000-0003-2448-1550},
B.~Adeva$^{47}$\lhcborcid{0000-0001-9756-3712},
M.~Adinolfi$^{55}$\lhcborcid{0000-0002-1326-1264},
P.~Adlarson$^{87,42}$\lhcborcid{0000-0001-6280-3851}
% ...

{\footnotesize \it

% ...
$^{38}$Nikhef National Institute for Subatomic Physics, Amsterdam, Netherlands\\
% ...
$^{42}$National Center for Nuclear Research (NCBJ), Warsaw, Poland\\
% ...
$^{47}$Instituto Galego de F{\'\i}sica de Altas Enerx{\'\i}as (IGFAE), Universidade de Santiago de Compostela, Santiago de Compostela, Spain\\
% ...
$^{51}$Physik-Institut, Universit{\"a}t Z{\"u}rich, Z{\"u}rich, Switzerland\\
% ...
$^{55}$H.H. Wills Physics Laboratory, University of Bristol, Bristol, United Kingdom\\
% ...
$^{57}$Department of Physics, University of Warwick, Coventry, United Kingdom\\
% ...
$^{59}$School of Physics and Astronomy, University of Edinburgh, Edinburgh, United Kingdom\\
$^{60}$School of Physics and Astronomy, University of Glasgow, Glasgow, United Kingdom\\
$^{61}$Oliver Lodge Laboratory, University of Liverpool, Liverpool, United Kingdom\\
% ...
$^{69}$Syracuse University, Syracuse, NY, United States\\
% ...
$^{87}$Department of Physics and Astronomy, Uppsala University, Uppsala, Sweden, associated to $^{60}$\\
% ...
}
\end{flushleft}